\documentclass[twocolumn,showpacs,prb]{revtex4}

\usepackage{epsfig}
\usepackage{dcolumn}
\usepackage{bm}

\newcolumntype{.}{D{.}{.}{-1}} 

\begin{document}

\title{First-principles study of superconductivity in
high-pressure Boron}
\author{D. A. Papaconstantopoulos}
 \email{papacon@dave.nrl.navy.mil}
\author{M. J. Mehl}
 \email{mehl@dave.nrl.navy.mil}
\affiliation{Center for Computational Materials Science,
Naval Research Laboratory, Washington, D.C. 20375-5345}

\date{\today}

\begin{abstract}
We study superconductivity in Boron using first-principles LAPW
calculations, the rigid-muffin-tin approximation and the McMillan
theory.  Our results point to an electron-phonon mechanism producing
transition temperatures near 10K at high pressures in agreement with
recent measurements.
\end{abstract}



\maketitle

\section{\label{sec:intro}Introduction}

In a recent paper Eremets \emph{et al.}\cite{eremets01:boron}
reported experiments where boron transforms from a non-metal at
normal pressures to a superconductor at very high pressures above
160 GPa.  They presented results that showed an increase of the
superconducting temperature, $T_c$, from 6K at 175~GPa to 11K at
250~GPa.

Previous theoretical studies on the metallization of boron based on
first-principles total-energy calculations\cite{mailhiot90:boron}
predicted that the 12 atom insulating form of boron ($\alpha_{12}$B)
at atmospheric pressure transformed to a metallic body-centered
tetragonal phase at 210 GPa and subsequently to an fcc structure at
360 GPa.

In this work we performed total energy and band structure
calculations for the rhombohedral $\alpha_{12}$B and fcc phases.
Our results for the fcc phase were used as an input to the
rigid-muffin-tin (RMT) theory of Gaspari and
Gyorffy\cite{gaspari72:lambda} to determine the value of the
Hopfield parameter $\eta$ and subsequently evaluate the coupling
constant $\lambda$ and $T_c$.  Our results for the parameter $\eta$
give values comparable to those reported in the
past\cite{papacon77:h} for metallic hydrogen at very high pressure.
Assuming that the fcc phase is the real high pressure structure in
the experiments of Eremets \emph{et al.}\cite{eremets01:boron}, this
indicates that the simple RMT theory accounts for high-pressure
superconductivity in boron.

\section{\label{sec:work}Theory and Results}

In this work we have applied the RMT theory\cite{gaspari72:lambda}
to calculate the Hopfield parameter $\eta$ given by the expression:

\begin{equation}
\eta = \frac{E_F}{\pi^2 N(E_F)} \sum_\ell 2(\ell+1)
\sin^2(\delta_{\ell+1}-\delta_\ell) \frac{N_\ell
N_{\ell+1}}{N_\ell^{(1)} N_{\ell+1}^{(1)}}
\label{equ:hopfield}
\end{equation}
where $\delta_l$ is the scattering phase shift at the Fermi energy
$E_F$ and angular momentum $\ell$, $N_{\ell}^{(1)}$ is the
single-scatterer density of states which, as defined in
Ref.~\onlinecite{gaspari72:lambda}, is an integral involving the
radial wave functions. $N(E_F)$ is the total density of states (DOS)
at $E_F$ and $N_\ell$ are the angular momentum components of the DOS
inside the muffin-tin spheres.  Eq.~(\ref{equ:hopfield}) is exact to
$\ell = 1$, but for higher values of $\ell$ it involves
non-spherical corrections.  The necessary input to
Eq.~(\ref{equ:hopfield}) was generated from a set of full potential
Linearized Augmented Plane Wave (LAPW) calculations that we
performed for fcc-B using touching muffin-tin spheres, and the
tetrahedron method for the DOS.

\begin{figure}
\includegraphics[width=3.4in]{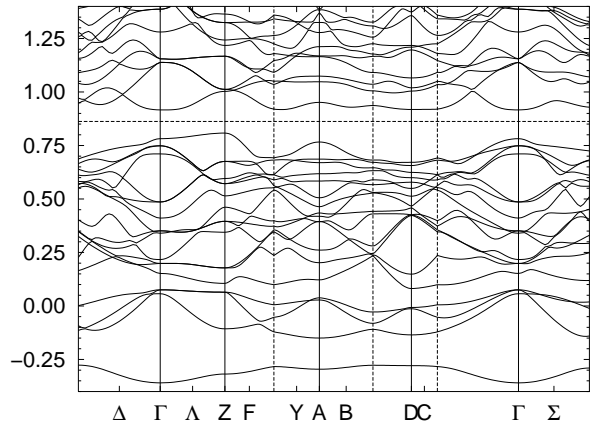}
\caption{\label{fig:alphaband}The electronic band structure of
$\alpha_{12}$B, calculated by the LAPW method.  The solid vertical
lines represent high-symmetry points, while the dashed vertical
lines represent zone boundaries.  The horizontal line near 0.85 Ry
is the Fermi level.}
\end{figure}

\begin{figure}
\includegraphics[width=3.4in]{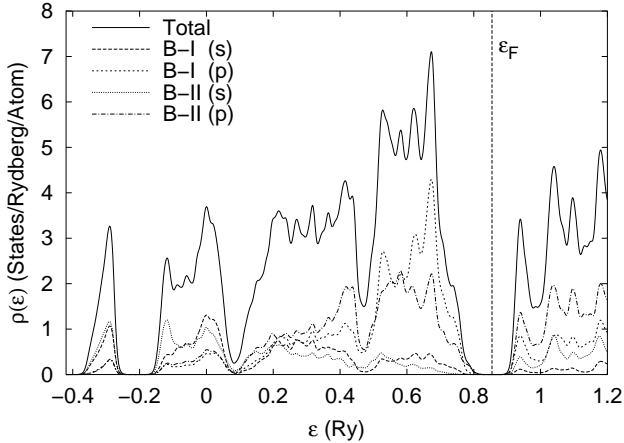}
\caption{\label{fig:alphados}The electronic density of states of
$\alpha_{12}$B, computed by smearing the eigenvalues of an LAPW
calculation using a Fermi temperature of 5~meV.  The $s$ and $p$
decomposition is shown for both types of atoms (B-I and B-II) in the
$\alpha_{12}$B structure.}
\end{figure}

\begin{figure}
\includegraphics[width=3.4in]{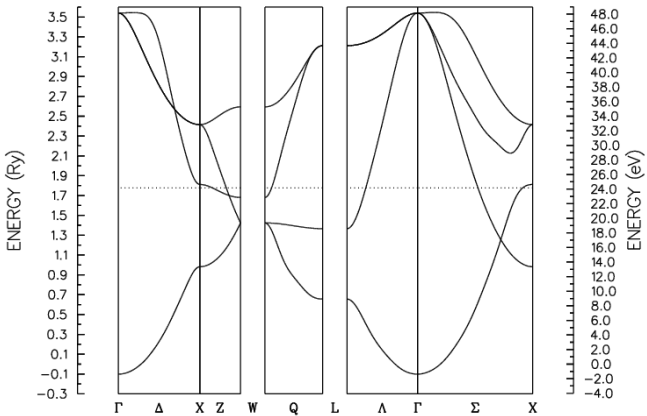}
\caption{\label{fig:fccband}The electronic band structure of fcc B
at the lattice constant 4.60~a.u.}
\end{figure}

We also performed an LAPW calculation for $\alpha_{12}$B.This is a
rhombohedral structure, space group R$\overline{3}$m-D$_{3d}^5$
(\#166), with measured lattice parameters\cite{pearson_handbook} $a
= b = c = 9.56$~a.u. and angles $\alpha = \beta = \gamma =
58.06^\circ$.  The energy bands of $\alpha_{12}$B are shown in
Fig.~\ref{fig:alphaband}.  We obtain a valence bandwidth of 15.9~eV
and an energy gap of 1.47~eV separating valence and conduction
bands.  The angular momentum character of the states can be seen
from Fig.~\ref{fig:alphados}, which shows the expected dominance of
the $p$-states through the whole spectrum.

For fcc B we calculated an equilibrium lattice parameter
a=5.37~a.u. and a bulk modulus of 282~GPa, to be compared with the
values of 5.34~a.u. and 269~GPa reported by Mailhiot \emph{et
al.}\cite{mailhiot90:boron} We note that for $a = 4.60$~a.u., which,
according to our total energy calculations corresponds to a pressure
of 307~GPa, the nearest neighbor B-B distance is 3.25~a.u., which is
close to the B-B distance (3.37~a.u.) in the newly discovered
superconductor MgB$_2$.  This leads us to believe that
superconductivity may have the same origin in both materials.

\begin{figure}
\includegraphics[width=3.4in]{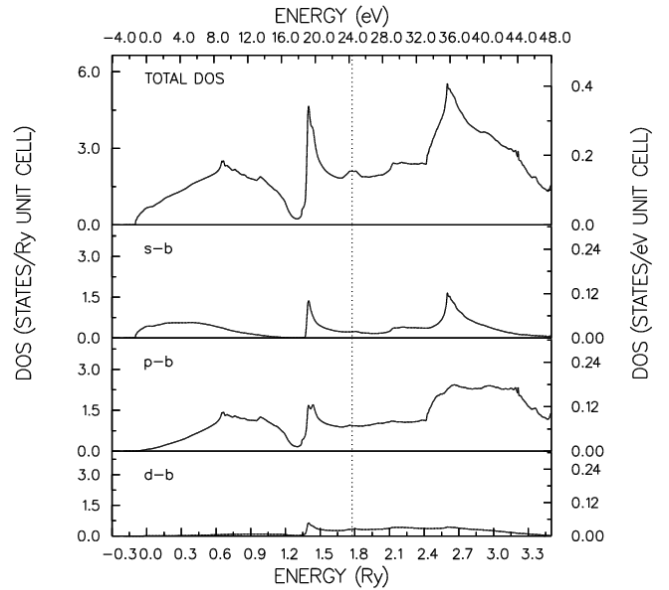}
\caption{\label{fig:fccdos}The electronic density of states of fcc B
at the lattice constant 4.60~a.u., as determined by the tetrahedron
method.}
\end{figure}

The band structure of fcc B confirms a metallic phase for all
lattice parameters, with a rapidly increasing band width going to
higher pressures.  The occupied band with is 1.43~Ry at equilibrium
and 1.88~Ry at $a = 4.60$~a.u.  The energy bands for $a =
4.60$~a.u. are shown in Fig.~\ref{fig:fccband}.  A fairly flat band
appears near $E_F$ in the XW direction at all lattice parameters.
At X the band has strong $s$ character, becoming $p$-like at W.
This situation is reminiscent of MgB$_2$ where a similar flat band
just above $E_F$ has been identified as the origin of
superconductivity in this
compound.\cite{an01:_super,kortus01:_super,mehl01:_cub2,kong01:_elect}
This band appears to be responsible for a peak in the DOS at $E_F$
shown in Fig.~\ref{fig:fccdos}.  Also in Fig.~\ref{fig:fccdos} we
find the $s$, $p$, and $d$ distribution of the DOS.  In
Fig.~\ref{fig:fccdos} we confirm that the strongest DOS component is
that from $p$-states.  The values of DOS at $E_F$ enter
Eq.~(\ref{equ:hopfield}) above for the evaluation of the Hopfield
parameter $\eta$.  The input and results using
Eq.~(\ref{equ:hopfield}) at four lattice parameters are listed in
Table~\ref{tab:eta}.

\begin{table*}
\caption{\label{tab:eta}Radius of the muffin-tin sphere $R_s$, Fermi
level $E_F$,Total DOS at $E_F$, angular components of the DOS
$N_\ell$, free-scatterer DOS $N_\ell^{(1)}$, scattering phase shifts
$\delta(\ell)$, and $\ell$ components of the Hopfield parameter
$\eta$ and the total value of $\eta$.  The three columns are headed
by the lattice parameters and the corresponding pressures.}
\begin{ruledtabular}
\begin{tabular}{l....}
a (a.u.) & 4.60  & 5.00  & 5.37  & 6.00  \\
P (GPa) & \multicolumn{1}{c}{307} & \multicolumn{1}{c}{89} &
\multicolumn{1}{c}{0} & \multicolumn{1}{c}{-50} \\
\hline
$R_s$ (a.u.)           & 1.626 &  1.768 & 1.898 & 2.121 \\
$E_F$ (Ry)  	       & 1.776 &  1.408 & 1.139 & 0.784 \\
$N(E_F)$ (states/Ry/spin)   & 1.050 & 1.328 & 1.475 & 1.514 \\
$N_s(E_F)$ (states/Ry/spin) & 0.114 & 0.172 & 0.233 & 0.276 \\
$N_p(E_F)$ (states/Ry/spin) & 0.469 & 0.568 & 0.637 & 0.769 \\
$N_d(E_F)$ (states/Ry/spin) & 0.164 &  0.193 & 0.195 & 0.155 \\
$N_s^{(1)}$   (states/Ry/spin) & 0.187 &  0.231 & 0.270 & 0.324 \\
$N_p^{(1)}$   (states/Ry/spin) & 0.818 &  1.058 & 1.334 & 1.991 \\
$N_d^{(1)}$   (states/Ry/spin) & 0.010 &  0.105 & 0.106 & 0.099 \\
$\delta_s$    & -0.109 &  0.099 & 0.298 & 0.658 \\
$\delta_p$    &  0.682 &  0.761 & 0.835 & 0.962 \\
$\delta_d$    &  0.031 &  0.029 & 0.027 & 0.021 \\
$\eta_{sp}$ (eV/\AA$^{2}$) &  2.945 &  1.576 & 0.820 & 0.150 \\
$\eta_{pd}$ (eV/\AA$^{2}$) & 11.555 &  9.172 & 6.960 & 4.048 \\
$\eta_{df}$ (eV/\AA$^{2}$) &  0.089 &  0.063 & 0.044 & 0.022 \\
$\eta_{tot}$ (eV/\AA$^{2}$)& 14.588 & 10.811 & 7.824 & 4.220 \\
\end{tabular}
\end{ruledtabular}
\end{table*}

We note that in Table~\ref{tab:eta} the $\ell$-components of the DOS
are given within the MT spheres (as is the case in
Fig.~\ref{fig:fccdos}),and therefore they do not add up to the value
of $N(E_F)$. The p-component, $N_p(E_F)$, is, as expected, dominant
and is about 60\% of the value of the total inside the MT.  The p-d
scattering, resulting from the $\ell=1$ term in
Eq.~(\ref{equ:hopfield}), gives by far the largest contribution to
the Hopfield parameter $\eta$. For the high pressure case ($a =
4.60$~a.u.) $\eta_{pd} = 0.79 \times \eta_{tot}$. The value of
$\eta_{tot}$ is approximately the same as the value of $\eta =
14.13$~eV/\AA$^2$ previously reported\cite{papacon77:h} for metallic
hydrogen at a higher pressure of 467~GPa.  It is also interesting to
note that the boron value of $\eta$ at equilibrium is very close to
the value $\eta = 7.627$~eV/\AA$^2$ for Nb,\cite{papacon77:super} a
typical transition metal superconductor where the d-f scattering is
the dominant term in Eq.~(\ref{equ:hopfield}).

To evaluate the critical temperature $T_c$ we have used the
McMillan\cite{mcmillan68:tc} approach defining an electron-phonon
coupling constant $\lambda =
\eta/M<\hspace{-4pt}\omega^2\hspace{-4pt}>$.  Here $\eta$ is
calculated from Eq.~(\ref{equ:hopfield}) as discussed above. For the
average phonon frequency $<\hspace{-4pt}\omega\hspace{-4pt}>$ we
followed Eremets \emph{et al.}\cite{eremets01:boron} and chose a
range of values from 1200K to 1400K.  The resulting $\lambda$ is in
the range 0.53 to 0.39 respectively.  Continuing we use the McMillan
equation:
\begin{equation}
T_c = \frac{<\hspace{-4pt}\omega\hspace{-4pt}>}{1.45} \exp \left[-
\frac{1.04(1+\lambda)}{\lambda -\mu^* (1+0.62\lambda)}\right] ~ .
\label{equ:mcmil}
\end{equation}

\begin{figure}
\includegraphics[width=3.4in]{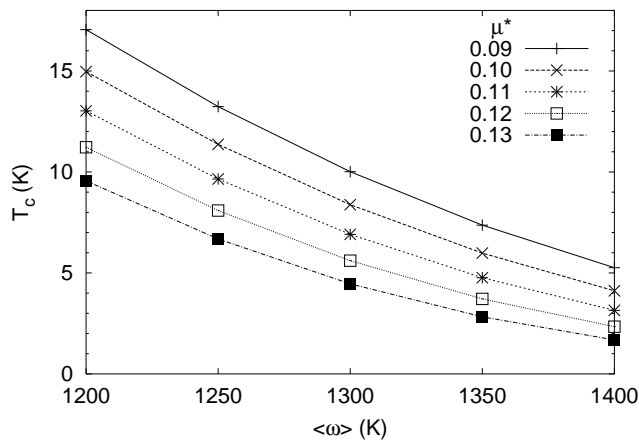}
\caption{\label{fig:tc}The superconducting transition temperature
$T_c$ of fcc Boron, as determined by the McMillan equation
(\protect\ref{equ:mcmil}), as a function of the RMS phonon frequency
$<\hspace{-4pt}\omega\hspace{-4pt}>$ at various values of
$\mu^*$.}
\end{figure}

We solve Eq.~(\ref{equ:mcmil}) for five values of the Coulomb
pseudopotential $\mu^*$ = 0.09 to 0.13, and in the above range of
$\omega$ and $\lambda$ values.  The corresponding values of $T_c$
are shown in Fig.~\ref{fig:tc}.  Within the uncertainty of the
values of the frequency $\omega$, and near the value $\omega =
1250$K our model predicts $T_c$ values in the range of the
experiment which validates the assertion of an electron-phonon
mechanism.

Finally, we want to comment on the pressure dependence of $T_c$.
Our calculations show that the parameter $\eta$ goes from the value
7.82~eV/\AA$^2$ at the equilibrium fcc volume to the value of
14.59~eV/\AA$^2$ at a pressure of 307~GPa.  It is important to note
that $N(E_F)$ follows the opposite trend, i.e., it reduces with
increasing pressure.  Assuming that the variation of
$<\hspace{-4pt}\omega\hspace{-4pt}>$ with pressure is not as strong
as the variation of $\eta$ we propose that the rapid increase of
$\eta$ is responsible for the observed increase of Tc with pressure.

In summary, we find that LAPW calculations, the RMT theory, and a
McMillan analysis for $T_c$ give a good description of
superconductivity in boron at high pressures.

\begin{acknowledgments}
This work was supported by the U.S. Office of Naval Research.
\end{acknowledgments}


\end{document}